# Investigation of all niobium Nano-SQUIDs based on sub-micrometer cross-type Josephson junctions

M Schmelz[1*], Y Matsui[2], R Stolz[1], V Zakosarenko[3], T Schönau[1], S Anders[1], S Linzen[1], H Itozaki[2], and H-G Meyer[1]

[1] Leibniz Institute of Photonic Technology, Albert-Einstein-Straße 9, D-07745 Jena, Germany
[2] Graduate School of Engineering Science, Osaka University, Osaka 560-8531, Japan
[3] Supracon AG, An der Lehmgrube 11, D-07751 Jena, Germany

* Corresponding author. Tel.: +49 3641 206122; fax: +49 3641 206199
  E-mail address: matthias.schmelz@ipht-jena.de

**Abstract**

We report on the development of highly sensitive SQUIDs featuring sub-micrometer loop dimensions. The integration of high quality and low capacitance SIS $Nb/AlO_x/Nb$ cross-type Josephson tunnel junctions results in white flux noise levels as low as 66 $n\Phi_0/Hz^{1/2}$, well below state-of-the-art values of their Nb-based counterparts based on constriction type junctions. Estimation of the spin sensitivity of the best SQUIDs yield $S_\mu^{1/2} < 7\mu_B/Hz^{1/2}$ in the white noise region, suitable for the investigation of small spin systems.
We discuss fabrication challenges, show results on the electrical characterization of devices with various pickup loops, and describe options for further improvement, which may push the sensitivity of such devices even to single spin resolution.

______________________________________________________________________

## 1. Introduction

In recent years there has been growing interest in the investigation of small spin systems like magnetic nanoparticles [1], molecular magnets [2] or single electrons and atoms [3] as well as in scanning SQUID microscopy [4]. For sensing magnetic fields on microscopic scales there are several sensors available, like e.g. Hall sensors [5], magnetic resonance force sensors [6], a method based on nitrogen vacancy defects in diamond [7], or SQUID type sensors [8]. Although the latter are known to be one of the most sensitive devices for measuring magnetic flux, they usually have dimensions of several μm to mm and are therefore not well adapted to the task of measuring small spin systems.

In order to improve the spin sensitivity $S_\mu^{1/2} = S_\Phi^{1/2}/\Phi_\mu$ of SQUIDs (here $S_\mu$ and $S_\Phi$ are the noise spectral power density normalized to a magnetic moment and flux, respectively), one needs to reduce their physical dimensions – thereby reducing the equivalent flux noise spectral density $S_\Phi$ via the decrease in total SQUID inductance $L_{SQ}$ [9], as well as increasing the coupling $\Phi_\mu$ between a particle with magnetic moment $\mu$ to the SQUID [10, 11]. Therefore in recent years, many attempts have been made for the development of miniature or even nanometer-sized SQUID sensors that may achieve the required sensitivities.

Such miniaturized SQUIDs are usually realized using constriction type junctions. Here, typically a small hole is patterned into a thin superconducting strip either by electron beam or focused ion beam lithography [12-15]. Some of these devices yield very low white flux noise levels of down to 0.2 $\mu\Phi_0/Hz^{1/2}$. More recently, very impressive results have been achieved by depositing a SQUID loop on the apex of a hollow quartz tube pulled into a very sharp pipette [16], resulting in white flux noise levels of down to 50 n$\Phi_0$/Hz$^{1/2}$ for Pb based devices. However, such sensors often show hysteretic current-voltage characteristics or have a very limited temperature working range for optimum performance, and have – depending on the material – a very limited lifetime of hours at ambient conditions. Moreover, the noise behavior is up to now not well understood in these devices, which makes their optimization difficult.

Another drawback of this approach is that the magnetic particle has to be placed close to the constriction for optimum coupling. This does not allow an independent optimization of the junction or SQUID properties and the coupling factor [17].

Unfortunately, the minimum dimensions of conventional window-type SIS Josephson junctions of a few micrometer [18, 19] up to now do not allow their implementation in SQUIDs with sub-micrometer loop dimensions. Even if their size could be reduced further, the parasitic capacitance due to the overlap of metallization layers around the junction will become more significant.

Therefore, the development of Nb/HfTi/Nb SNS sandwich-type junctions have been reported recently [20]. This group implemented Josephson tunnel junctions into nanometer-sized SQUID loops. Although the white flux noise levels are only a few 100 n$\Phi_0$/Hz$^{1/2}$, the performance probably suffers from a lower characteristic voltage than comparable SIS Josephson junctions.

In order to overcome the above mentioned drawbacks, we recently introduced a technology for the fabrication of miniature SQUIDs based on cross-type SIS Nb/AlO$_x$/Nb Josephson tunnel junctions [21]. First results on such devices with loop dimensions in the micrometer range were very promising and already exhibit white flux noise levels comparable to their much smaller state-of the-art counterparts based on constriction type junctions [11].

In this paper we report on further downsizing such devices to sub-micrometer SQUID loop dimensions. Details on the sample preparation, including current limitations, will be described in Section 2. We will show that the decrease in SQUID inductance results in a considerable decrease of white flux noise levels well below state-of-the-art values of Nb based devices. The measurement results will be discussed in Section 3 and conclusions will be drawn how to further improve sensor characteristics.

## 2. Sample Preparation

The devices described herein were fabricated in an extended version of our recently presented cross-type technology [21]. In addition to the process steps described therein, two thermally evaporated SiO isolation layers (each with a thickness of 400 nm) with a subsequent 300 nm Nb wiring layer are deposited, which allows the fabrication of SQUID sensors with on-chip integrated input coils on top of the SQUID washers. Furthermore, a layer of 100 nm thick AuPd is sputter deposited to electrically shunt the junctions. A final SiO layer with a thickness of 400 nm serves as protection layer on top.

One of the main advantages of this technology is the negligible parasitic capacitance from surroundings of the junctions, which together with their high quality enables the fabrication of low noise SQUID sensors, as already shown for SQUID magnetometers [22, 23] or current sensors [24]. Moreover, the technology allows the fabrication of homogenous sensor arrays, which may be of particular importance in e.g. SQUID microscopy or for the investigation of crystals of magnetic molecular clusters. Our investigation on series arrays of 100 unshunted Josephson junctions reveals a standard deviation of the critical current of about $1\sigma \approx 0.6$ % for $(3 \times 3)$ µm$^2$ junctions. It is further worth to note that the presented approach offers the possibility to place a magnetic particle near a local constriction in the SQUID loop, which allows further optimization of the coupling independent of the junction properties used for the optimization of SQUIDs.

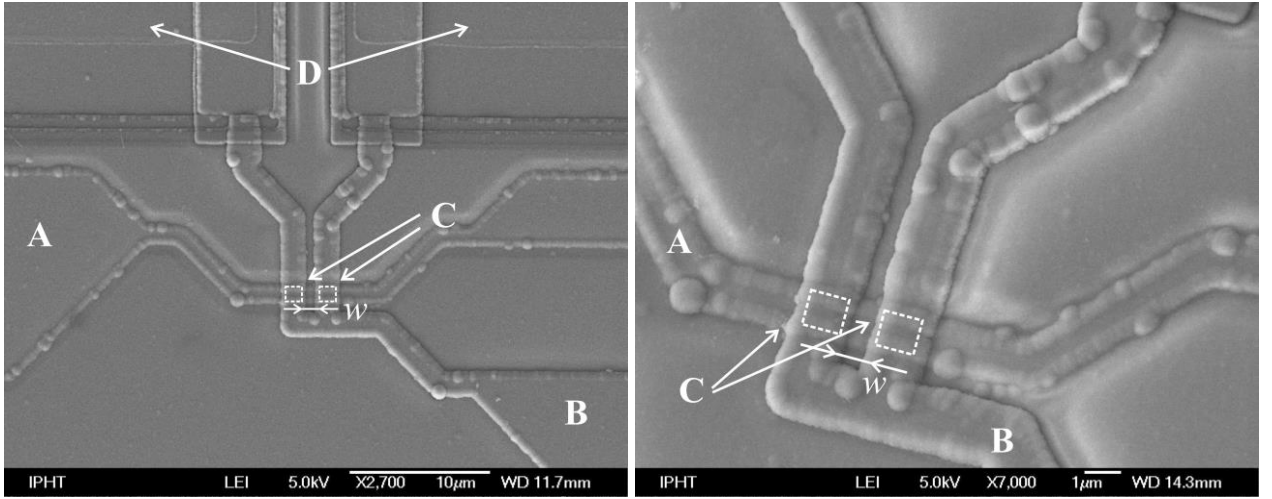

**Figure 1.** Scanning electron microscope images of a SQUID with $(0.8 \times 0.8)$ µm$^2$ Josephson junctions and an inner loop dimension $w$ of 1.5 µm. The Josephson junctions (C) are built by the overlap of trilayer (A) and Nb wiring layer (B). They are shunted by resistors implemented in a AuPd layer (D). Please note that visible structures are slightly enlarged due to overlying isolation layers with a total thickness of about 1200 nm.

The SQUIDs are formed by a narrow trilayer strip and a u-shape of the Nb wiring layer, where the two ends of the u-shape cross the perpendicular trilayer strip. Figure 1 shows a scanning electron microscope image of such a device with $(0.8 \times 0.8)$ µm$^2$ Josephson junctions and an inner loop diameter of $w = 1.5$ µm. The Josephson junctions are formed by the overlaps between the trilayer and the Nb wiring layer, as indicated by the two dotted squares.

Table 1 lists the dimensions of devices under investigation. The inner loop diameter $w$ was varied from 10 µm down to 0.5 µm, which represents the resolution of the used i-line stepper lithography tool. We typically achieve alignment errors between these two layers of less than 100 nm. The square shaped Nb/AlO$_x$/Nb Josephson junctions have a critical current density of 2.3 kA/cm$^2$ and linear dimensions of about 0.8 µm. According to Fiske-step measurements on junctions with larger geometric dimensions, they exhibit a total junction capacitance of about $C \approx 40$ fF [25]. In addition, Table 1 shows the electrical parameters of the devices at 4.2 K: the critical current $2I_C$ of the SQUIDs, their shunt resistance $R/2$, voltage swing $V_{pp}$, and the calculated McCumber parameter $\beta_C = 2\pi I_C R^2 C/\Phi_0$ of about unity for these optimized sensors.

**Table 1.** Device parameters of the investigated SQUIDs; $w$ denotes the inner diameter of the SQUID loop, as indicated in Figure 1, $2I_C$ the critical current of the SQUID, $R/2$ the resistance of the SQUID in voltage mode and $V_{pp}$ the usable voltage swing as measured at 4.2 K.

| SQUID # | loop diameter $w$ | $2I_C$ [µA] | $R/2$ [Ω] | $V_{pp}$ [µV] | $\beta_C$ |
|---|---|---|---|---|---|
| 1 | 10 µm | 29.7 | 12.1 | 306 | 1.09 |
| 2 | 10 µm | 28.7 | 12.1 | 300 | 1.06 |
| 3 | 5.5 µm | 28.6 | 11.9 | 320 | 1.02 |
| 4 | 5.5 µm | 25.9 | 11.9 | 258 | 0.92 |
| 5 | 3.0 µm | 31.8 | 11.9 | 366 | 1.13 |
| 6 | 3.0 µm | 29.6 | 11.9 | 328 | 1.05 |
| 7 | 1.5 µm | 26.8 | 12.1 | 310 | 0.99 |
| 8 | 1.5 µm | 29.0 | 12.0 | 340 | 1.05 |
| 9 | 0.5 µm | 32.0 | 12.0 | 300 | 1.16 |
| 10 | 0.5 µm | 34.5 | 11.9 | 285 | 1.23 |

## 3. Device characterization and discussion of measurement results

The devices were characterized within a specially designed cryo-dipstick, which contains a superconducting solenoid made from NbTi wire. It operates in persistent-current mode in order to provide stable dc magnetic field to adjust the working point of the miniaturized SQUIDs. The solenoid was calibrated using a SQUID magnetometer with known effective area.

The measured effective areas of the devices at 4.2 K, as listed in Table 2, are determined from the coupling to the magnetic field of the solenoid. They agree reasonably well with the estimation using inner and outer diameter: $A_{eff} \approx w' \cdot (w' + 2 \cdot 0.8 \, \mu m)$, where $w' = w + 200$ nm represents the known change of design values during fabrication. This expression, except for the smallest devices #9 and #10, yields results that differ less than a few percent from the measured values. For these two SQUIDs, however, the estimations yield effective areas that are too large, which is probably caused by the fact that small resist residues remained in the corners of the u-shape so that the Nb is not completely etched away. The sample to sample variation in the effective area becomes larger for devices with smaller loop diameters, supporting this hypothesis.

Since the investigated sensors are expected to exhibit very low white flux noise levels in the range of about 100 n$\Phi_0$/Hz$^{1/2}$, we used a two-stage read-out configuration for the measurement of the equivalent flux noise, as shown in Figure 2. Even with state-of-the-art low-noise directly-coupled SQUID electronics the measured noise would otherwise be dominated by the electronics.

In this setup, the signal from SQUID SQ$_1$ under investigation is fed to a superconducting input coil of a SQUID array SQ$_2$ via resistor $R_C$. The value of $R_C$ was about 0.5 $\Omega$, so the SQUID SQ$_1$ operated practically with voltage bias and the array SQ$_2$ serves as an ammeter. The series array consists of 16 SQUIDs and exhibits an input current noise of below 2.5 pA/Hz$^{1/2}$. Both SQUIDs were placed inside a $\mu$-metal and superconducting shields and were immersed in liquid helium at 4.2 K. SQUID SQ$_1$ was biased with flux and current so that the current swing sensed in the SQUID array due to a signal to SQ$_1$ was maximum. Feedback from a commercial low-noise directly-coupled flux locked loop electronics [26] was applied to SQ$_2$. The input signal of SQUID SQ$_1$ as well as a dc magnetic field to adjust the working point was provided via the solenoid. The SQUID array was placed at a small distance so that the influence of the magnetic field was negligible. From the output voltage $V_{out}$, recorded with an HP 3565 spectrum analyzer with a maximum bandwidth of 100 kHz, the equivalent flux noise of SQUID SQ$_1$ was calculated via the experimentally determined transfer function in each working point. Table 2 lists the measured white flux noise levels.

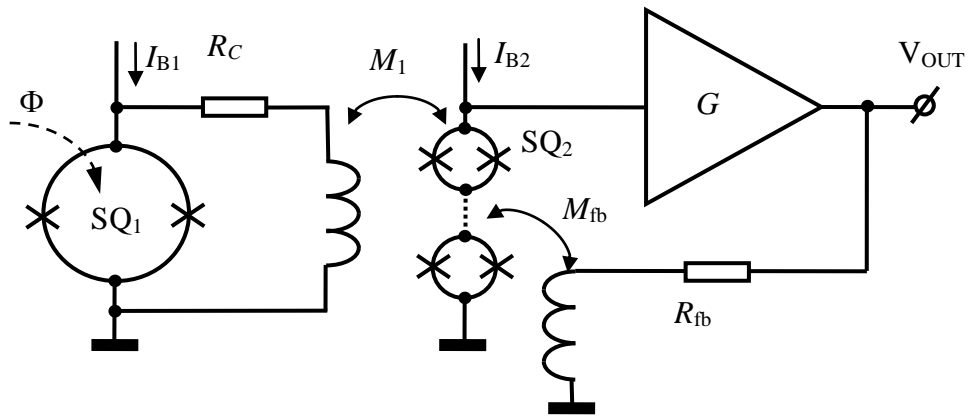

**Figure 2.** Schematic of the two-stage setup used for noise measurements. The flux noise of SQUID SQ$_1$ is measured with a second stage SQUID SQ$_2$ operated as a low noise preamplifier.

**Table 2.** Measured effective areas and equivalent white flux noise levels of the investigated SQUIDs at 4.2 K. Geometrical inductances result from FastHenry [27] simulations; estimated kinetic inductance values are based on $L_{kin} = \mu_0\lambda_L^2 \cdot s/(dh)$ [28], with $s$ as the circumference of the SQUID loop, $d$ and $h$ the width and height of the cross section of the SQUID loop, respectively. $\lambda_L$ is the London penetration depth and $\mu_0$ the vacuum permeability. Estimations of the white flux noise are according to relation (2) as explained in the text.

| SQUID # | inverse effective area [$\mu T/\Phi_0$] | geometrical inductance $L_{geo}$ [pH] | kinetic inductance $L_{kin}$ [pH] | $\beta_L$ | white flux noise [$n\Phi_0/Hz^{1/2}$] measured | estimated (2) |
|---|---|---|---|---|---|---|
| 1 | 15.8 | 32.3 | 8.60 | 0.61 | 230 | 207 |
| 2 | 15.7 | 32.3 | 8.60 | 0.59 | 200 | 207 |
| 3 | 45.7 | 16.4 | 5.32 | 0.31 | 160 | 119 |
| 4 | 45.4 | 16.4 | 5.32 | 0.28 | 125 | 119 |
| 5 | 128 | 8.62 | 3.50 | 0.19 | 127 | 89.1 |
| 6 | 130 | 8.62 | 3.50 | 0.18 | 106 | 89.1 |
| 7 | 381 | 4.52 | 2.41 | 0.09 | 90 | 67.4 |
| 8 | 402 | 4.52 | 2.41 | 0.10 | 76 | 67.4 |
| 9 | 2730 | 2.16 | 1.68 | 0.06 | 66 | 50.2 |
| 10 | 2280 | 2.16 | 1.68 | 0.07 | 66 | 50.2 |

The preamplifier noise contribution was already removed therein. For SQUID #9 this contribution amounts to 23.5 $n\Phi_0/Hz^{1/2}$, which was quadratically subtracted from the total measured noise of 70 $n\Phi_0/Hz^{1/2}$. Here and thereafter the measured noise spectral power density is normalized to the flux in SQUID using the measured maximal transfer function $\partial V/\partial \Phi = 10.3$ V/$\Phi_0$ for SQUID #9.

The upper (red) curve in Figure 3 shows the measured flux noise spectrum of SQUID #9 in an optimal (magnetic sensitive) working point with $\partial V/\partial \Phi > 0$. This sensor exhibits a very low white flux noise level of down to 66 n$\Phi_0$/Hz$^{1/2}$, well below state-of-the-art values of Nb-based counterparts with constriction type junctions. It is worth to point out, that the measured magnitude of flux noise at 1 Hz amounts to only 0.4 µ$\Phi_0$/Hz$^{1/2}$, corresponding to an energy resolution $\varepsilon = S_\Phi/(2L_{SQ})$ of about 3.4 $h$ in the white noise region and to 126 $h$ at 1 Hz, with $h$ being Planck's constant.

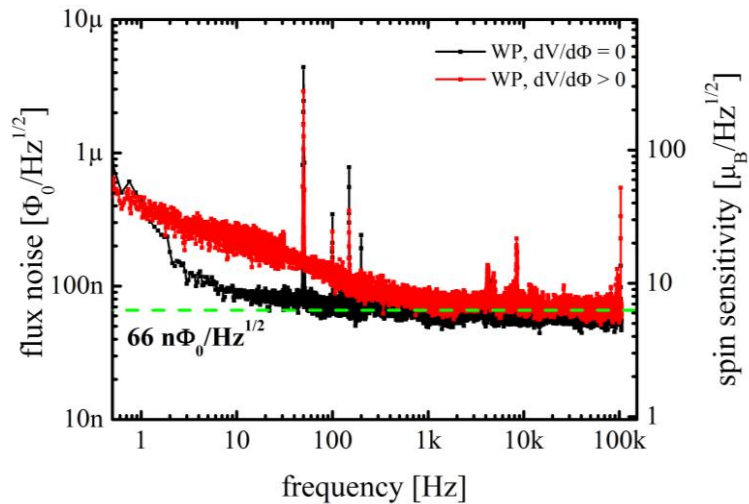

**Figure 3.** Equivalent flux noise spectrum for SQUID #9. The red and black curve correspond to working points (WP) on the slope of the flux-voltage characteristics with $\partial V/\partial \Phi > 0$ and $\partial V/\partial \Phi = 0$ where the SQUID is insensitive to magnetic flux noise, respectively. The equivalent white flux noise corresponds to 66 $n\Phi_0/Hz^{1/2}$. Note that the magnitude of flux noise at 1 Hz equals 0.4 $\mu\Phi_0/Hz^{1/2}$. The right hand axis was calculated according to $S_\mu^{1/2} = S_\Phi^{1/2}/\Phi_\mu$, with $\Phi_\mu = 10.5$ $n\Phi_0/\mu_B$, as explained in the text.

For the estimation of the total SQUID inductance $L_{SQ}$, we take the contribution from the kinetic inductance $L_{kin}$ into account, as the thickness of the superconducting layers in our devices are below the London penetration depth $\lambda_L$, which we assumed to be $\lambda_L = 90$ nm for our Nb films. Values for the geometrical inductances $L_{geo}$, as listed in Table 2, are based on simulations using FastHenry [27]. The kinetic inductance was estimated as $L_{kin} = \mu_0\lambda_L^2 \cdot s/(dh)$, with $s$ as the circumference of the SQUID loop, $d$ and $h$ the width and height of the cross section of the SQUID loop and $\mu_0$ the vacuum permeability [28]. The total SQUID inductance is thus given by $L_{SQ} = L_{geo} + L_{kin}$.

Measurements in a magnetic insensitive working point with $\partial V/\partial \Phi = 0$ (lower/ black curve in Figure 3) indicate that the increase of the low-frequency flux noise for 5 Hz $< f <$ 1 kHz is not due to critical current fluctuations in the Josephson junctions, but is caused by magnetic noise, in agreement with previous results obtained on SQUID magnetometers with much larger effective areas [22]. Since the dimensions of superconducting structures in the vicinity of the SQUID hole are very small, magnetic noise caused by the motion of vortices trapped in the SQUID washer is very unlikely in these devices and we currently attribute this noise to effects such as fluctuating spins at the surface of the superconductor [29] which has to be proved in future investigations.

The right hand axis in Figure 3 shows the calculated spin sensitivity $S_\mu^{1/2} = S_\Phi^{1/2}/\Phi_\mu$. Here, we estimated the coupling of a point dipole with magnetic moment of Bohr magneton $\mu_B$ in the center of a square SQUID washer with inner side length $2a = 0.5$ µm to $\Phi_\mu = \sqrt{2}/\pi \cdot \mu_0\mu_B/a = 10.5$ n$\Phi_0$/$\mu_B$ [11]. This results in a white spin sensitivity of better than 7 $\mu_B$/Hz$^{1/2}$.

In order to compare our measurement results to theoretical predictions of the white flux noise, we can use the widely known relation [30]
$$\varepsilon = 16\, k_BT\, (L_{SQ}C)^{1/2} \quad \text{and} \quad S_\Phi^{1/2} = 4\, L_{SQ}^{3/4}\, C^{1/4}\, (2k_BT)^{1/2}, \tag{1}$$
which is valid for optimized SQUIDs having $\beta_L = 2I_CL_{SQ}/\Phi_0 \approx 1$ and $\beta_C \approx 1$. However, as device dimensions and therefore the SQUID inductance $L_{SQ}$ vary by more than an order of magnitude for the developed sensors, the screening factor $\beta_L$ does not meet this requirement (cf. Table 2). Nonetheless, for SQUIDs with $w = 10$ µm the measured flux noise is only slightly higher than this theoretical prediction. Moreover, deviations between measured and calculated values increase with decreasing $\beta_L$, indicating the limit of the above relation. Besides it suggests a way for further improvements: increasing $\beta_L$ to unity by increasing the junction's critical current may allow even lower white flux noise values.

In contrast to the above relation, we suggest discussing the noise behavior of the investigated SQUIDs as following: Expression (1) can be treated as the thermal energy $k_BT$ distributed in the frequency range limited by the SQUID time constant $\tau_{LC} = (L_{SQ}C/2)^{1/2}$. If the SQUID is not optimized, we have to use the longest time constant limiting the SQUID bandwidth, which is $\tau = RC \approx 1$ps in our SQUIDs. Substituting $(L_{SQ}C/2)^{1/2} = \tau = RC$ in (1) for SQUIDs with non-optimized screening factor $\beta_L$ yields:
$$\varepsilon = 16\sqrt{2}\, k_BT\, (RC)^{1/2} \quad \text{and} \quad S_\Phi^{1/2} = 8\, (L_{SQ}RC\, k_BT)^{1/2}. \tag{2}$$
Here we use the total SQUID inductance $L_{SQ} = L_{geo}+L_{kin}$ because both parts of inductance contribute to the SQUID energy. The resulting values of white flux noise levels estimated from (2) are given in Table 2.

One can see that the measured noise is typically a factor of about 1.3 larger than the estimated values independent of $\beta_L$. This may allow for more reliable estimations of noise figures for future Nano-SQUID sensor designs.

## 4. Conclusion

In conclusion, we have described ongoing work towards highly sensitive nanoSQUIDs based on cross-type Josephson tunnel junctions. We demonstrated the integration of these high quality and low capacitance SIS junctions in SQUIDs featuring sub-micrometer loop dimensions. Reducing device dimensions results in a substantial decrease in white flux noise levels. The smallest devices exhibit white flux noise levels as low as 66 n$\Phi_0$/Hz$^{1/2}$, well below state-of-the-art values of Nb-based counterparts with constriction type junctions.

Optimizing the dimensional screening parameter $\beta_L$ to about unity still leaves room for further improvements. The future challenges include the increase of the critical current density of the junctions, without – if possible – impairing the low-frequency noise performance of the sensors. Besides a further reduction in SQUID loop dimensions, the optimization of the coupling of a magnetic particle to the SQUID by e.g. local narrow constrictions in the SQUID loop may push the sensitivity of such devices even to single spin resolution.

Compared to the typically used Nano-SQUIDs based on constriction type junctions, the possibility to reliably fabricate homogenous sensor arrays represents a unique feature of our approach. Moreover, these SQUIDs should offer a continuous operation over a broad temperature range down to mK, essential for the investigation of magnetization dynamics.

## References

1. Wernsdorfer W, Classical and Quantum Magnetization Reversal Studied in Nanometer-Sized Particles and Clusters. In *Advances in Chemical Physics*, John Wiley & Sons, Inc.: 2001; pp 99-190.
2. Wernsdorfer W 2007 *Nat Mater* **6** 174-176
3. Bushev P, Bothner D, Nagel J, Kemmler M, Konovalenko K B, Lörincz A, Ilin K, Siegel M, Koelle D, Kleiner R and Schmidt-Kaler F 2011 *Eur. Phys. J. D* **63** 9-16
4. Kirtley J R 2010 *Reports on Progress in Physics* **73** 126501
5. van Veen R G, Verbruggen A H, van der Drift E, Radelaar S, Anders S and Jaeger H M 1999 *Review of Scientific Instruments* **70** 1767
6. Rugar D, Budakian R, Mamin H J and Chui B W 2004 *Nature* **430** 329-332
7. Taylor J M, Cappellaro P, Childress L, Jiang L, Budker D, Hemmer P R, Yacoby A, Walsworth R and Lukin M D 2008 *Nat Phys* **4** 810-816
8. Clarke J and Braginski A I, *The SQUID Handbook: Fundamentals and Technology of SQUIDs and SQUID Systems*. Wiley-VCH: 2006.
9. Tesche C D and Clarke J 1977 *Journal of Low Temperature Physics* **29** 301-331
10. Ketchen M B, Awschalom D D, Gallagher W J, Kleinsasser A W, Sandstrom R L, Rozen J R and Bumble B 1989 *IEEE Transactions on Magnetics* **25** 1212-1215
11. Schmelz M, Stolz R, Zakosarenko V, Anders S, Fritzsch L, Roth H and Meyer H G 2012 *Physica C: Superconductivity* **476** 77-80
12. Hao L, Macfarlane J C, Gallop J C, Cox D, Beyer J, Drung D and Schurig T 2008 *Applied Physics Letters* **92** 192507
13. Lam S K H and Tilbrook D L 2003 *Applied Physics Letters* **82** 1078 - 1080
14. Troeman A G P, Derking H, Borger B, Pleikies J, Veldhuis D and Hilgenkamp H 2007 *Nano Letters* **7** 2152-2156
15. Hasselbach K, Veauvy C and Mailly D 2000 *Physica C: Superconductivity* **332** 140-147
16. Vasyukov D, Anahory Y, Embon L, Halbertal D, Cuppens J, Neeman L, Finkler A, Segev Y, Myasoedov Y, Rappaport M L, Huber M E and Zeldov E 2013 *Nat Nano* **8** 639-644

17. Wölbing R, Schwarz T, Müller B, Nagel J, Kemmler M, Kleiner R and Koelle D 2014 *arXiv: 1406.1402v2*
18. Stolz R, Fritzsch L and Meyer H G 1999 *Superconductor Science and Technology* **12** 806-808
19. Yohannes D, Sarwana S, Tolpygo S K, Sahu A, Polyakov Y A and Semenov V K 2005 *IEEE Transactions on Applied Superconductivity* **15** 90-93
20. Nagel J, Kieler O F, Weimann T, Wölbing R, Kohlmann J, Zorin A B, Kleiner R, Koelle D and Kemmler M 2011 *Applied Physics Letters* **99** 032506
21. Anders S, Schmelz M, Fritzsch L, Stolz R, Zakosarenko V, Schönau T and Meyer H G 2009 *Superconductor Science and Technology* **22** 064012
22. Schmelz M, Stolz R, Zakosarenko V, Schönau T, Anders S, Fritzsch L, Mück M and Meyer H G 2011 *Superconductor Science and Technology* **24** 065009
23. Schönau T, Schmelz M, Zakosarenko V, Stolz R, Meyer M, Anders S, Fritzsch L and Meyer H G 2013 *Superconductor Science and Technology* **26** 035013
24. Zakosarenko V, Schmelz M, Stolz R, Schönau T, Fritzsch L, Anders S and Meyer H G 2012 *Superconductor Science and Technology* **25** 095014
25. Schmelz M, Stolz R, Zakosarenko V, Anders S, Fritzsch L, Schubert M and Meyer H G 2011 *Superconductor Science and Technology* **24** 015005
26. Supracon AG, An der Lehmgrube 11, 07751 Jena, Germany. http://www.supracon.com/
27. Kamon M, Tsuk M J and White J K 1994 *IEEE Transactions on Microwave Theory and Techniques* **42** 1750-1758
28. Majer J B, Butcher J R and Mooij J E 2002 *Applied Physics Letters* **80** 3638-3640
29. Koch R, DiVincenzo D and Clarke J 2007 *Physical Review Letters* **98** 267003
30. Clarke J, SQUID Concepts and Systems. In *Superconducting Electronics*, Weinstock, H.; Nisenoff, M., Eds. Springer Verlag: Berlin/ Heidelberg/ New York, 1989; Vol. 59, pp 87-148.